\documentclass[10 pt, letter]{article}
\usepackage{graphicx}
\usepackage{subfigure}
\usepackage{caption}
\usepackage{amsmath}
\usepackage{amsthm}
\usepackage{enumerate}
\usepackage{subfig}
\usepackage{float}
\usepackage{amsfonts}
\usepackage{amssymb}
\usepackage{anysize}
\usepackage{helvet}
\usepackage{setspace}
\usepackage{multirow}
\usepackage{tabularx}
\usepackage{tabulary}
\usepackage{longtable}
\usepackage{float}
\usepackage{mathtools}
\usepackage[none]{hyphenat}

\usepackage{cite}
\usepackage[sort&compress,authoryear,comma,merge]{natbib}
\title{\bf Study of galaxy morphology and  merging time of two interacting galaxies under different initial rotation and orientation configurations}
\author{Elkin L. L\'opez\footnote{lopezlopezelkin@gmail.com}, Gustavo V. L\'opez\footnote{gulopez@cencar.udg.mx}, Simon N. Kemp \footnote{snk@astro.iam.udg.mx}.
	\\
	\\
 Departamento de F\'{i}sica, Universidad de Guadalajara,\\
 Blvd. Marcelino Garc\'{i}a Barragan y Calzada Ol\'{i}mpica, \\ CP 44200, Guadalajara, Jalisco, M\'exico, \\ 
 \\
$^{\ddagger}$Instituto de Astronomia y Meteorologia, Universidad de Guadalajara,\\ 
Av Vallarta 2602, Col Arcos Vallarta, \\
CP 44130, Guadalajara, Jalisco, M\'exico,  
 }
\begin{document}
\maketitle

\begin{abstract}
\noindent
Using the GADGET-2 N-body code, we make a study of the galaxy morphology and  merging time due to  two interacting galaxies (for the same types and different sizes and masses, $1:1$ and $1:10$ ratio masses) merging due to gravity interaction. This is done for different initial relative orientation and rotation of these galaxies (modes of interaction) but with the same relative bulge separation and the same relative initial velocities. It was found that the resulting galaxy morphology resemble many of the observed galaxies in our Universe, and that,  in general, a binary galaxy system with  1:10 mass ratio has larger merging time than  a binary galaxy system with  1:1 mass ratio. This difference is due to the different  evolution of the masses during the interaction in both cases. For the case with a 1:10 mass ratio, the global mass maximum is located at the end evolution, meaning that the second galaxy increases its mass constantly. For the case  with mass ratio 1:1, the global maximum is located around $t=0.35$ Gy, causing a reduction of the merging time.
\end{abstract}
\vskip1cm

\newpage
\section{Introduction}
The interaction of two galaxies is a very common process in our Universe \citep{lar}, and the observed galaxy morphology of the merger product depends on the time after the initial encounter in which it is observed \citep{boylan,mar}. At the end of the interaction we may have a `new' galaxy with different bulge and disk components and different distributions of gas, dust, and stars \citep{WuCox,MoCox,kar}. It is of interest to know the differences in the resulting galaxy morphology and the merging time for different initial orientations of both galaxies, the  size and mass of the resultant bulge/ellipsoid and disk components, and the different distributions of  gas and stars in the final merger product. \\ \\
\noindent
There have been many studies of simulations of the interaction of two galaxies \citep{mor,wal,pon}  mapping the exchange of gas and stars \citep{PhiCox,GuCox}. \cite{kar} studied the distribution of deposition of mass via minor mergers ($\mu = 0.1 - 0.25$) and mini mergers ($\mu = 0.01 - 0.1$) finding that mini mergers deposit a larger fraction of the stars in the exterior parts of the galaxy compared with minor mergers and do not contribute much to the central mass. Mini mergers can increase the size of the main disk component significantly. The find that streams result from circular satellite orbits while shells form from more radial orbits. \cite{kim} investigate the formation of warped disks of spiral galaxies in `fly-by' encounters of these galaxies with adjacent dark matter haloes, finding that such encounters can produce warps that last several Gigayears.
 However, as far as we know, there has no been a systematic study of this interaction with all possible basic orientation of galaxies.  Here we present a systematic study of the galaxy morphology and merging time  of two interacting galaxies with different orientations of their rotation axes and different sense of rotation of the disk. We use the GADGET-2 N-body code \citep{gadget2,gadget2-2,gadget2-3} considering different relative orientations and rotation. The galaxies are initially at the same separation and have the same relative velocity and we start the orbit of the second galaxy in the tangential direction.\\

\section{Configurations} 
We consider two symmetrical fried-egg-shaped galaxies with bulge \citep{bouok,milion,bertin,launhardt,dwek,xu,salviander}
and disk  components \citep{vanture,bertin,giallongo,kennicutt,kenni2,burkert} containing stars, gas, and dust 
\citep{bertin,kacprzak,ivison,daylan}, plus a dark matter halo \citep{bode,oppen}. The centre of the bulge defines the trajectory of the galaxy \citep{holtzman,magorrian}, which usually coincides with the centre of the dark matter halo \citep{risa}. Galaxies exist in binary systems and in small groups (less than 100 members) or large clusters of galaxies \citep{turnary,pauling,voit, gallaghere}. \\ \\
\noindent
We will adopt the following convention for the initial configuration of the two galaxies with respect to the Cartesian reference system: (a) If the plane of the galaxy's disk is in the x-y plane of the system and it is rotating in a counter-clockwise direction, we denote it as $\textrm{Z}^{+}$ (axis of rotation). If this galaxy is rotating in a clockwise direction, we denote it as $\textrm{Z}^{-}$. (b) If the plane of the galaxy's  is in the x-z plane of the reference system and  is rotating in a counter-clockwise direction, we denote it as $\textrm{Y}^{+}$ (axis of rotation), and for its clockwise direction as $\textrm{Y}^{-}$. (c) If theplane of the  galaxy's disk  is in the z-y plane of the reference system, we denote it as $\textrm{X}^{+}$  for the counter-clockwise direction of rotation, and $\textrm{X}^{-}$ for the clockwise  direction. For each galaxy we therefore consider  six cases, and so we have  a total of 36 possible combinations of orientation and rotation  of the two galaxies in the interaction: $(\textrm{Z}^{+},\textrm{Z}^{+}),~ (\textrm{Z}^{+},\textrm{X}^{+}),~\dots ~, (\textrm{Z}^{-},\textrm{X}^{-})$. In addition, we consider two  mass relations: (a) when the central (primary) galaxy is ten times the mass of the orbiting (secondary) galaxy, (b) when they are of equal masses. To denote the interaction configuration of the pair we then use a pair of terms,  the first term is the primary galaxy's rotation state, and the second term is the secondary galaxy's rotation state.\\ 
\\
Figure 1 shows the initial positions for $(\mbox{Z}^+,\mbox{X}^+)$ and $(\mbox{Z}^+,\mbox{Y}^+)$, having different masses. In all cases under our consideration, the galaxies will have the same distance separation at the beginning of the simulation.
\begin{figure}[H]
\includegraphics[width=\textwidth]{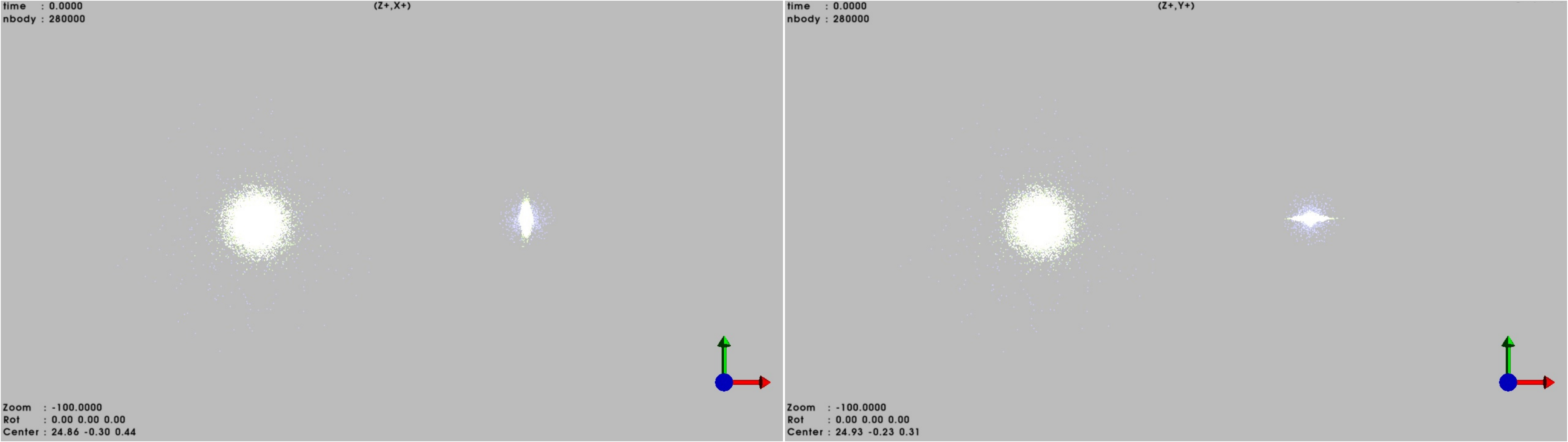}
\centering
    \caption{Two interacting galaxies with different modes; $(\mbox{Z}^+,\mbox{X}^+)$ on the left and $(\mbox{Z}^+,\mbox{Y}^+)$ on the right}
\end{figure}
\section{Simulations}
We use our available version of the GADGET-2  N-body code to make the simulations of the interactions mentioned above. Details of this code can be found in the above mentioned references, but we want to state some  important characteristics used in our simulations.  If   $r_{200}$ is the radius containing three-quarters of the galaxy mass, where the density is 200 times the critical density of the universe and $v_{200}$ is the rotation velocity or velocity dispersion at this radius, then the total mass of the galaxy within this radius  is  $M_{200}=v_{200}^2 r_{200}/ G$, where $G$ is the gravitational constant ($G=6.674\times 10^{-11}$ m$^3$kg$^{-1}$s$^{-2}$, \cite{rosetion}).
\begin{equation}
M_{200}=M_{bulge}+M_{halos}+M_{disk}.
\end{equation}
The halo and bulge densities, as a function of the radius ($r$) of the galaxy, are 
\begin{equation}
\rho_{halo,bulge}(r)=\frac{M_{halo, bulge}}{2\pi}\frac{a_{h,b}}{r(r+a_{h,b})^3},
\end{equation}
where the parameter $a_{h,b}$,  in terms of the concentration index $c$,  is $a_h=r_s\sqrt{2[\ln(1+c)-c/(1+c)}$, and 
the halo scale length is defined as $r_s=r_{200}/c$ and  the proportion bulge scale length per scale radius $f_b=a_b/R_d$, where $R_d$ is the scale length of the disk. The angular momentum of the halo is 
\begin{equation}
J=\lambda\sqrt{2GM_{200}^3r_{200}/f_c},
\end{equation}
where $\lambda$ is the twist parameter, and the parameter $f_c$, written in terms of the concentration index, is $f_c=c[1-1/(1+c)^2-2\ln(1+c)/(1-c)]/(2[\ln(1+c)-c/(1+c)]^2)$. The fractional angular momentum of the bulge is  $J_b=JM_{bulge}/M_{200}$, and the angular momentum of the disk is  $J_d=j_dJ$, $j_d$ being a free parameter.  The disk (stellar) density varies as
\begin{equation}
\label{dendisk}
\rho_{*}(R,z)=\frac{M_{*}}{4\pi z_0 R_d^2}sech^2\left(\frac{z}{z_0}\right)\exp \left( -\frac{R}{R_d}\right),
\end{equation}
where $R=\sqrt{x^2+y^2}$, $z_0$ is the parameter which determines the thickness of the disk, and $M_*$ is the stellar mass on the disk. Similarly, the gas in the disk has a surface density of
\begin{equation}
\Sigma_{gas}=\frac{M_{gas}}{2\pi h_r^2}\exp \left(-\dfrac{r}{h_r} \right),
\end{equation} 
where $h_r$ is the scale length of the gas profile, and $M_{gas}+M_{*}=M_{disk}$. 
In addition, the vertical structure of gas in asymmetric galaxies is governed by the equation
\begin{equation}
\frac{1}{\rho_{gas}}\frac{\partial P}{\partial z}+\frac{\partial \Phi}{\partial z}=0,
\end{equation} 
where $\Phi$ is the gravitational potential due to the total mass of the gas.\\ 
\\
Our simulation of two interacting galaxies with the GADGET-2 code were performed with the following set of parameters  for the primary galaxy  $G_1$ and the secondary galaxy $G_2$ ($N_i$ being the number of elements of each part of the galaxy):
\begin{flushleft}
	\begin{tabular}{l l c r r}
		${\bf G_1}:$ & & & &
		\\
		$~~~c=10.0$, & $v_{200}=120~\mathrm{km/s}$,& $R_d=1.04722$ h$^{-1}$kpc, & $\lambda =0.033$, & $M_{d}/M_{200}=0.04$,\\
		$~~~M_{b}/M_{200}=0.09$, & $M_{gas}/M_{d}=0.1$, & $a_b/R_d=0.2$, & $z_0/R_d=0.2$, & $J_{d}/J=0.04$, \\
		$~~~N_{halo}=150,000$, & $N_{disk}=20,000$, & $N_{gas}=20,000$, & $N_{bulge}=10,000$ .& \\
		& & & & \\
		${\bf G_2}:$ & & & &\\
		$~~~c=10.0$, & $v_{200}=55.699~\mathrm{km/s}$, & $R_d=0.486075$ h$^{-1}$kpc & $\lambda =0.033$, & $M_{d}/M_{200}=0.04$,\\
		$~~~M_{b}/M_{200}=0.09$, & $M_{gas}/M_{d}=0.1$, & $a_b/R_d=0.2$, & $z_0/R_d=0.1$, & $J_{d}/J=0.04$, \\
		$~~~N_{halo}=30,000$, & $N_{disk}=20,000$, & $N_{gas}=20,000$, & $N_{bulge}=10,000$. & \\
				& & & & \\
	\end{tabular}\\
\end{flushleft}
The initial conditions of our two galaxies for all modes of interaction  are: $\vec{x}_1=\vec{0}~\mbox{h}^{-1}$kpc, $\vec{x}_2=50\hat{x}~\mbox{h}^{-1}$kpc ($\mbox{h}= H_0/100$ km s$^{-1}$ Mpc$^{-1}$ observationally $\mbox{h}\sim 0.7$ and 1 kpc $ \approx 206264806~$AU with $1$ AU $\approx 149.6\times10^6 ~$km), the velocities are $\vec{v}_1=\vec{0}~\mathrm{km/s}$ and $\vec{v}_2=99.0548172\hat{y}~\mathrm{km/s}$. The subindex 1 represents the primary galaxy and 2 represents the secondary one. The baryonic masses of the galaxies are $\mathcal{M}_1=57.4\times 10^{10}$M$_\odot$ and $\mathcal{M}_2/\mathcal{M}_1=0.1,1$ (for 1:10 and 1:1 mass ratios respectively).\\
\\
We define the merging time (illustrated in Figure 2 and 9) as the time where the asymptotic behavior of the galaxies' separation and oscillations in separation reach defined limits, as a consequence of the fusion of the bulges. This time is calculated in the following way: given a small $\epsilon\in\Re$, we can find a time $t_{\epsilon}$, such that for any $t>t_{\epsilon}$, we have $\left| \dot{d}(t>t_\epsilon)\right|/\left| d(0)/ \Delta t\right| \le \epsilon$, and at the time $t_{\epsilon}$ we have $\left|\dot{d}(t_\epsilon)\right|/\left|d(0)/ \Delta t\right|> \epsilon$. 
Thus, time $t_{\epsilon}$ marks the division between oscillations of high amplitude ( $t \le t_{\epsilon}$) and oscillations of small amplitude ($t>t_{\epsilon}$). The merging time $\tau$ is defined as $\tau=t_\epsilon+q\Delta t$, where $q$ is the smallest positive integer that satisfies that $\tau$ is the closest time to $t_{\epsilon}$ that satisfies $\left| \dot{d}(\tau)\right|/\left| d(0)/ \Delta t \right| \approx 0$. For this estimation we used the following parameters: $\epsilon=0.004$ and $d(0)/ \Delta t=5000$ kms$^{-1}$ with $\Delta t=0.01$ (0.98 h$^{-1}$Gyr).\\
\\
The resulting galaxy morphology of the two galaxies interaction at the final  time $t_f$ of our simulation will be characterized  by the resulting parameters $M_{*}(t_f)$, $R_d(t_f)$, and $z_0(t_f)$. Although we also present the shape of the merged galaxy in different planes at different times. 
\section{Results}
\subsection{Galaxies interaction for $M_2=M_1/10$}
Figure 2 shows the evolution of the distance between the  centres of the galaxies' bulges  for all 36 interaction modes (rotation, orientation) of the galaxies with mass ratio 1:10. There are  six groups of 6 modes each (each group for a color), and in each group the primary galaxy has the same rotation and orientation mode. Each group has similar evolution of the separation during the interaction, indicating the behaviour is dominated by the the primary galaxy. \\\\
\noindent
As  imentioned above, the merging time is defined as when the oscillations in separation of the centres reach a determined minimum level (when the velocity derivative of the curve tends to zero). The merging time $\tau$ of the galaxies is shown in the second column of Table 1 for different masses (first number). 
The merging time for different masses is quicker for the $\mbox{Z}^{+}$ cases and slowest for the $\mbox{Z}^{-}$ cases. For $\mbox{Z}^{+}$ the primary galaxy is rotating in the same direction and in the same orbital plane as the orbital movement of the secondary galaxy, meaning that the stars are in closer contact for a longer time and so have more time to interact. The $\mbox{Z}^{-}$ cases have the longest merger time because the primary galaxy is rotating in the opposite direction to the orbital motion of the secondary galaxy. The rotating direction of the secondary galaxy has relatively little effect on the merging times.
\begin{figure}[H]
	\centering
	\centering
	\includegraphics[width=\textwidth]{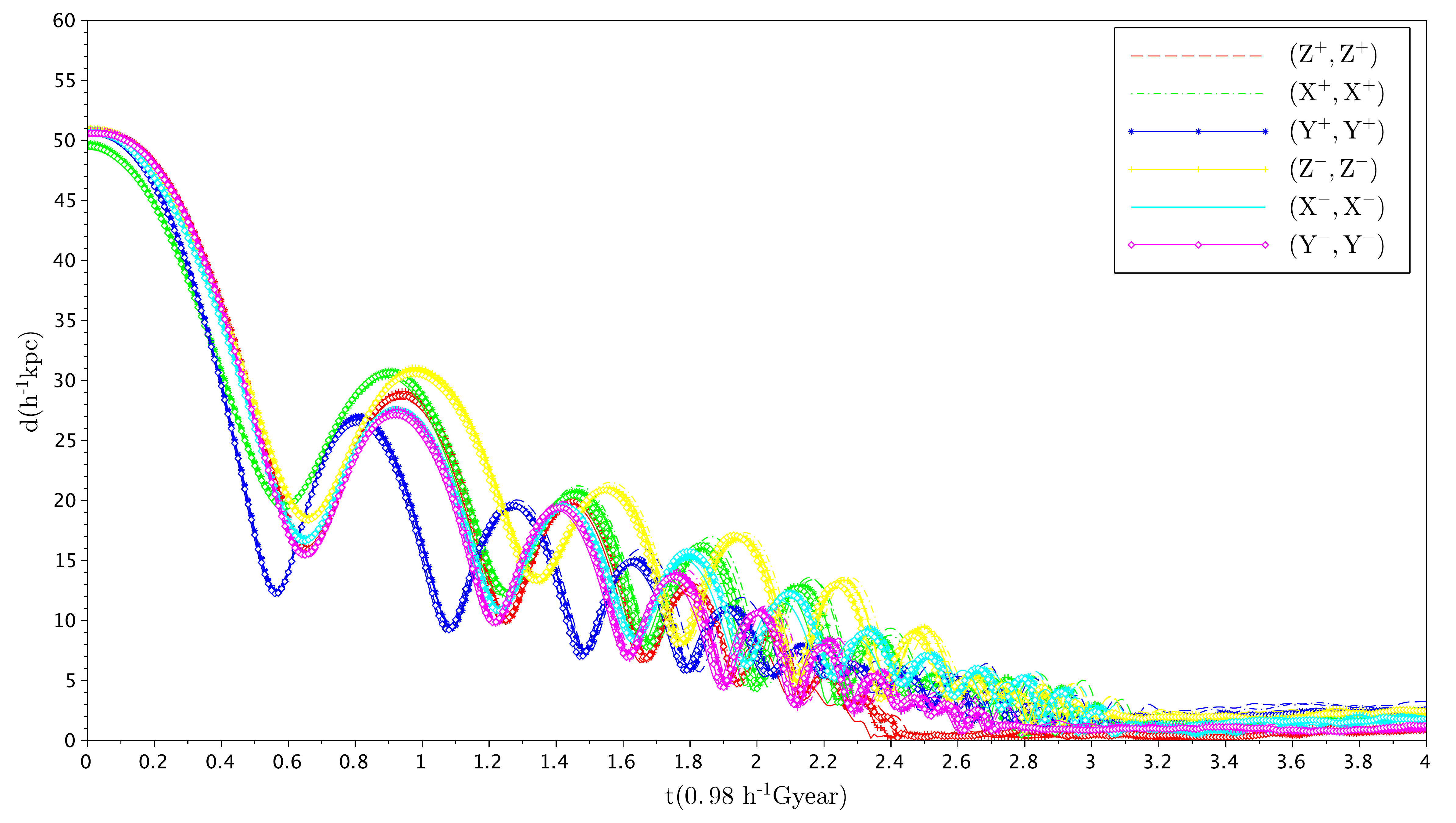}
    \caption{Time evolution of the separation of the centres of the  bulges  of two interacting galaxies for 36 modes.}
\end{figure}
\noindent
Figures 3--8 show the galaxiy morphologies and position of stars of the two galaxies, at time $t=0.0, 2.0, 4.0$ (0.98 h$^{-1}$Gyr), for all rotation and orientation interaction modes (36 modes) and for the mass ratio 1:10. These modes of interaction take a long time, and even at 4 Gyr the merged galaxy looks like a disk galaxy with bulge and, often, with a distorted antisymmetrical disk. This disk is frequently extended in agreement with  \cite{kar}. At the time 2 Gyr, there are structures which look like arcs in shell galaxies but in the plane of the disk, and their formation from a companion with tangential orbits, in contradicts the results of  \cite{kar}. 
We note that this interaction configuration favours the formation of dsk galaxies as a final merger product, as the tangential orbits of the secondary galaxy gives a high initial angular momentum to the system, which tends to remian in the disk component of the final product.

 \begin{figure}[H]
	\centering
	\includegraphics[page=1,height=0.45\textheight]{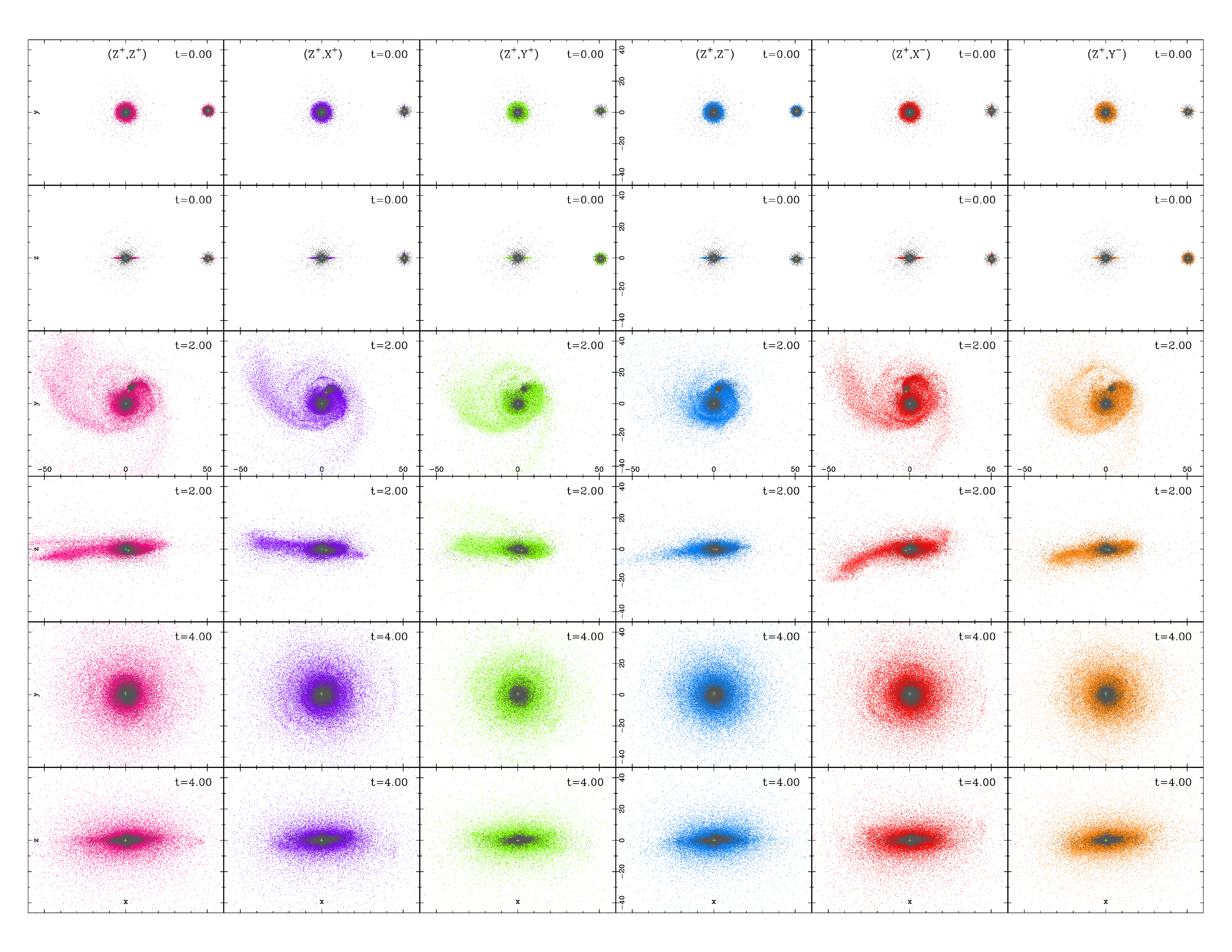}
		\caption{\small{Bulge stars in grey and disk stars coloured, for all interactions of mass ratio 1:10 with mode $\mbox{Z}^+$, each column showing one mode of interaction. The initial configuration $t=0.0$ Gyr (top), $t=2.0$ Gyr (middle), $t=4.0$ Gyr (bottom) . Each time has two perspectives, the x-y plane and x-z plane, one above the other.}}
	\includegraphics[page=2,height=0.45\textheight]{dif4}
	\caption{\small{As Fig. 3 for all  interactions with mass ratio 1:10 in which rotation of the primary galaxy is $\mbox{X}^+$}}
\end{figure}
\begin{figure}[H]
	\centering
	\includegraphics[page=3, height=0.45\textheight]{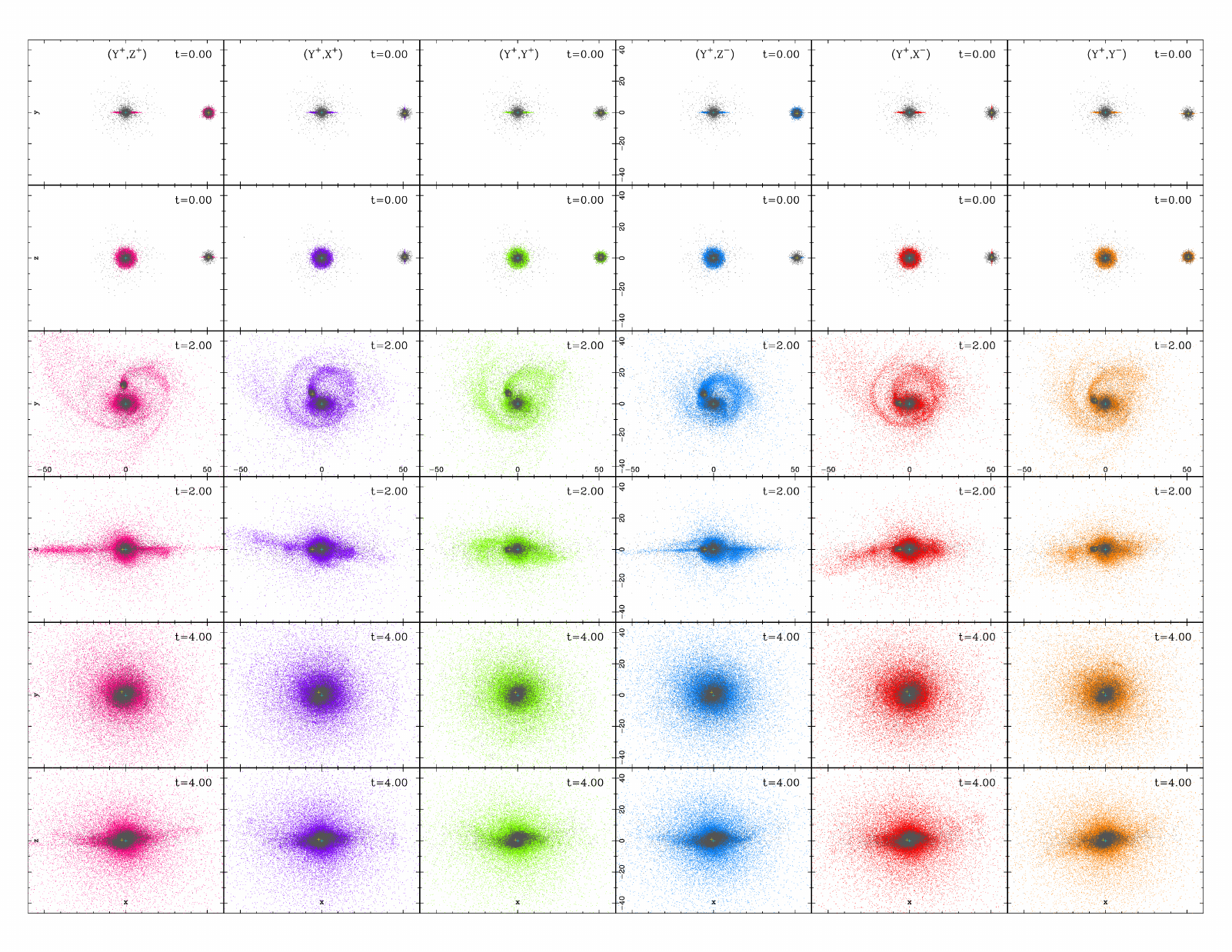}
		\caption{\small{As Fig. 3  for all interactions with mass ratio 1:10  in  which the rotation  of primary galaxy is $\mbox{Y}^+$}}
	\includegraphics[page=4, height=0.45\textheight]{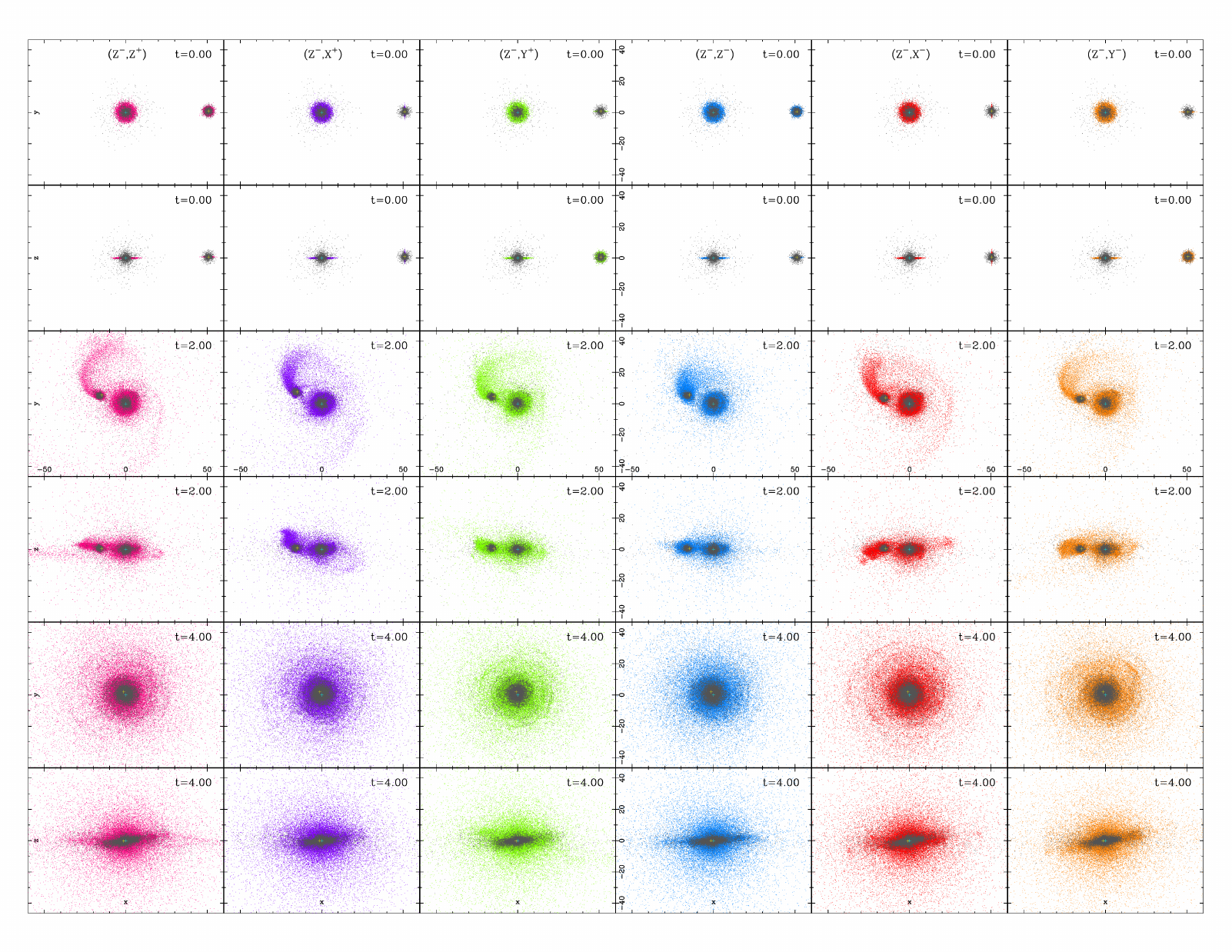}
		\caption{\small{ As Fig. 3 for all  interactions with mass ratio 1:10 in which rotation of the primary galaxy is $\mbox{Z}^-$.}}
\end{figure}
\begin{figure}[H]
	\centering
	\includegraphics[page=5, height=0.45\textheight]{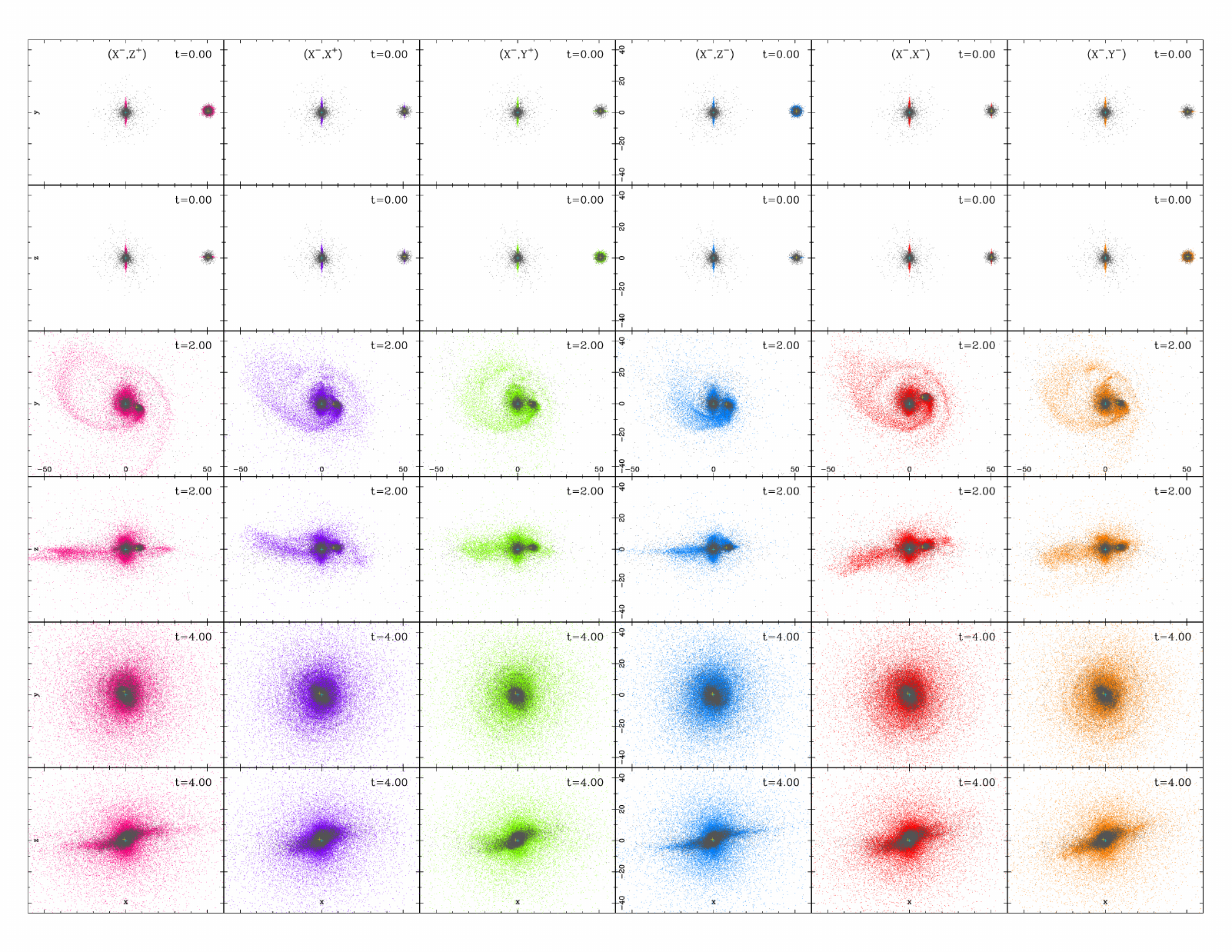}
		\caption{\small{As Fig. 3  for all interactions with mass ratio 1:10  in  which the rotation  of primary galaxy is $\mbox{X}^-$}}
	\includegraphics[page=6, height=0.45\textheight]{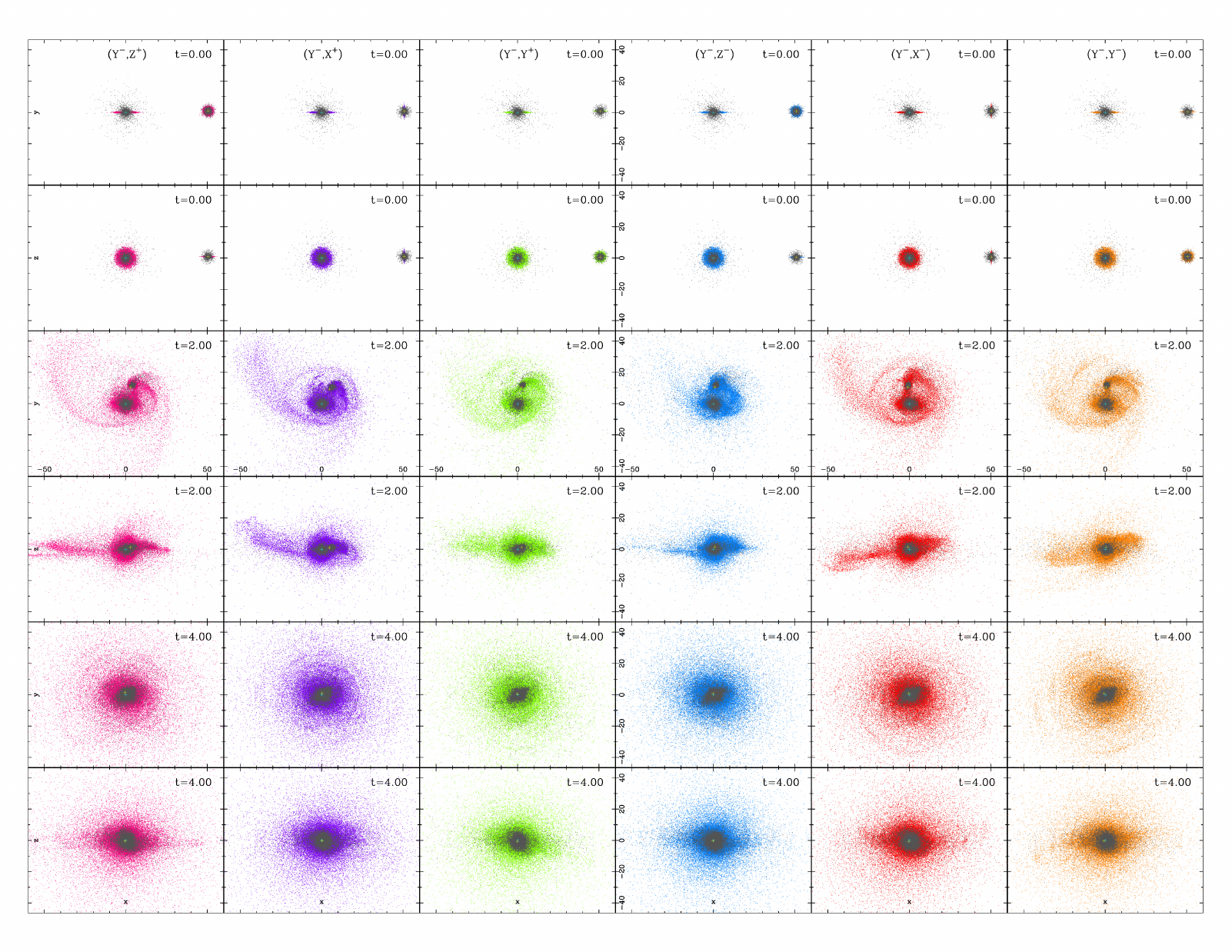}
		\caption{\small{ As Fig. 3  for all interactions with mass ratio 1:10  in  which the rotation  of primary galaxy is $\mbox{Y}^-$.}}
 \end{figure}
\subsection{Galaxies interaction for $M_2=M_1$}
We consider now that the second galaxy has the same parameters as the first one, and we repeat all the cases we had before. Figure 9 shows the evolution of distance of separation of the galaxies' bulges for all interaction modes. The merging time $\tau$ is shown in the second column (second number) of Table 1.   With the secondary galaxy having much more mass, the fusion times are much shorter, between 0.5--0.7 (0.98 h$^{-1}$Gyr), depending on if they are considered to be merged on the second or third close encounter.\\
\\
\begin{figure}[H]
	\centering
	\includegraphics[width=0.9\textwidth]{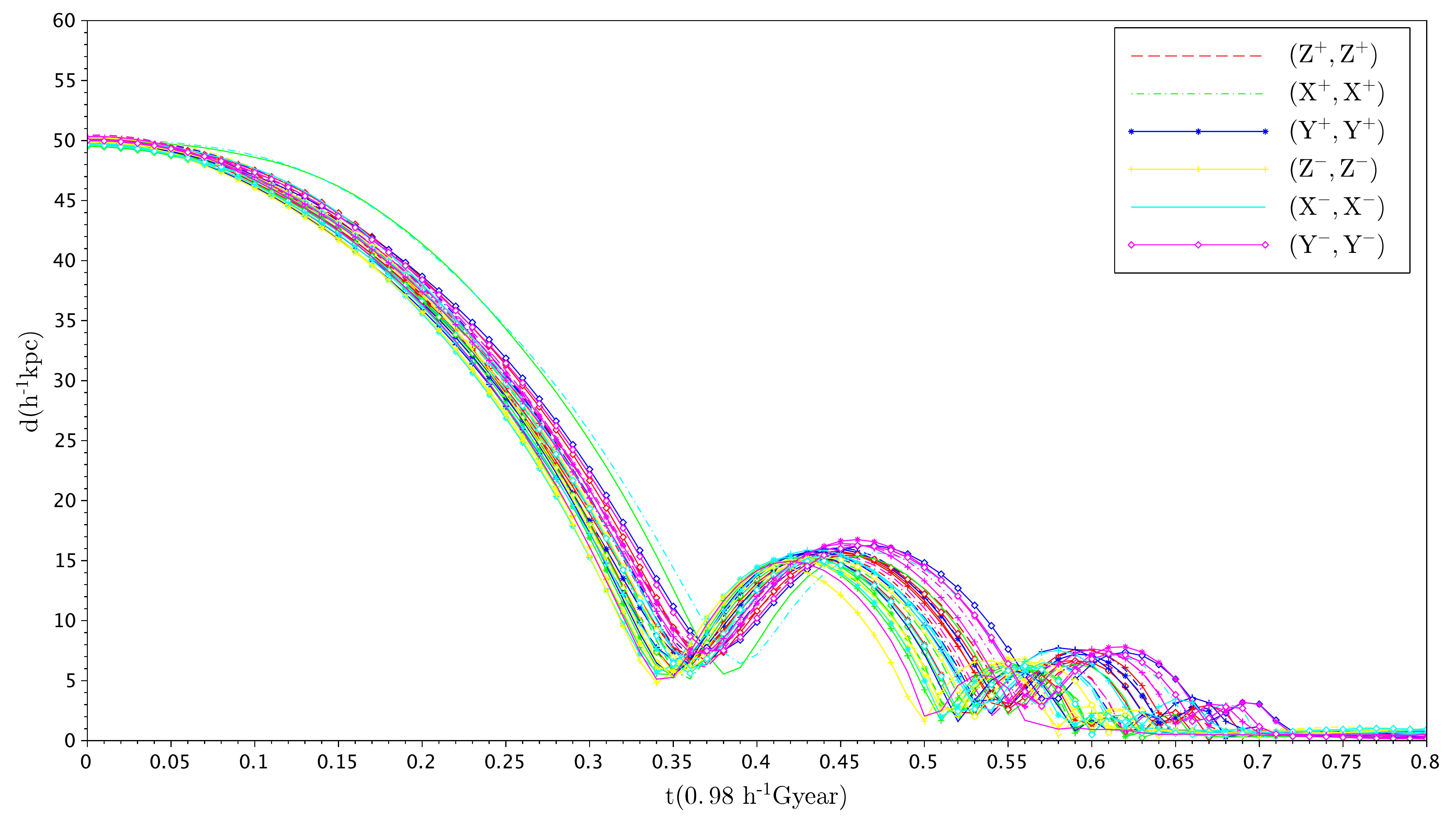}
        \caption{Time evolution of the separation of the bulges  of two interacting galaxies for 36 modes. }
\end{figure}
\noindent
Figures 10-15 show the galaxy morphologies and the positions of galaxy stars at times $t=0.0, 0.5, 1.0$ (0.98 h$^{-1}$Gyr), top to bottom in each column, for all rotation interaction modes (36 modes). The time intervals between frames are shorter because the merging process is more rapid in this case. Here the intermediate and final-stage disks are not as extended as in the 1:10 mass ratio cases, as expected for equal-mass mergers \cite{kar} and the arc-like structures in the disk plane in the  final stage look more like sections of spiral arms than shell structures. The cases with the secondary galaxy in the rotation state $\mbox{Z}^{+}$ all finish with a disturbed, asymmetric, warped disk, while $\mbox{Z}^{+}$,$\mbox{Z}^{+}$ is more regular with a box-peanut-like bulge. The other orientations seem more dominated by the galaxy morphology of the bulge/ellipsoid in the final stage, again some of these are box-peanut-like. In this case,with the secondary galaxy as massive as the first, the initial angular momentum of the system is even higher when the secondary galaxy is in the  $\mbox{Z}^{+}$ rotation state, favouring the formation of an extended disk structure. If the secondary galaxy is another rotation state the final merger product is more elliptical-like.\\
\\
Figure 16 shows a comparison of two different visualizations of the $(\mbox{X}^+,\mbox{Z}^-)$ interaction for mass ratio 1:10 at time $t=4.0$ (0.98 h$^{-1}$Gyr), the left hand side shows all baryonic galaxy stars and the right hand side shows only the disk stars of the galaxy. It can be seen that the initial disk stars spread out and some form part of the bulge of the merger product. Both images are similar but  there are more stars further from the plane above and below the bulge in the image with all galaxy stars. the same angled, rather distorted disk is seen in both.

\begin{figure}[H]
	\centering
	\includegraphics[page=1, height=0.45\textheight]{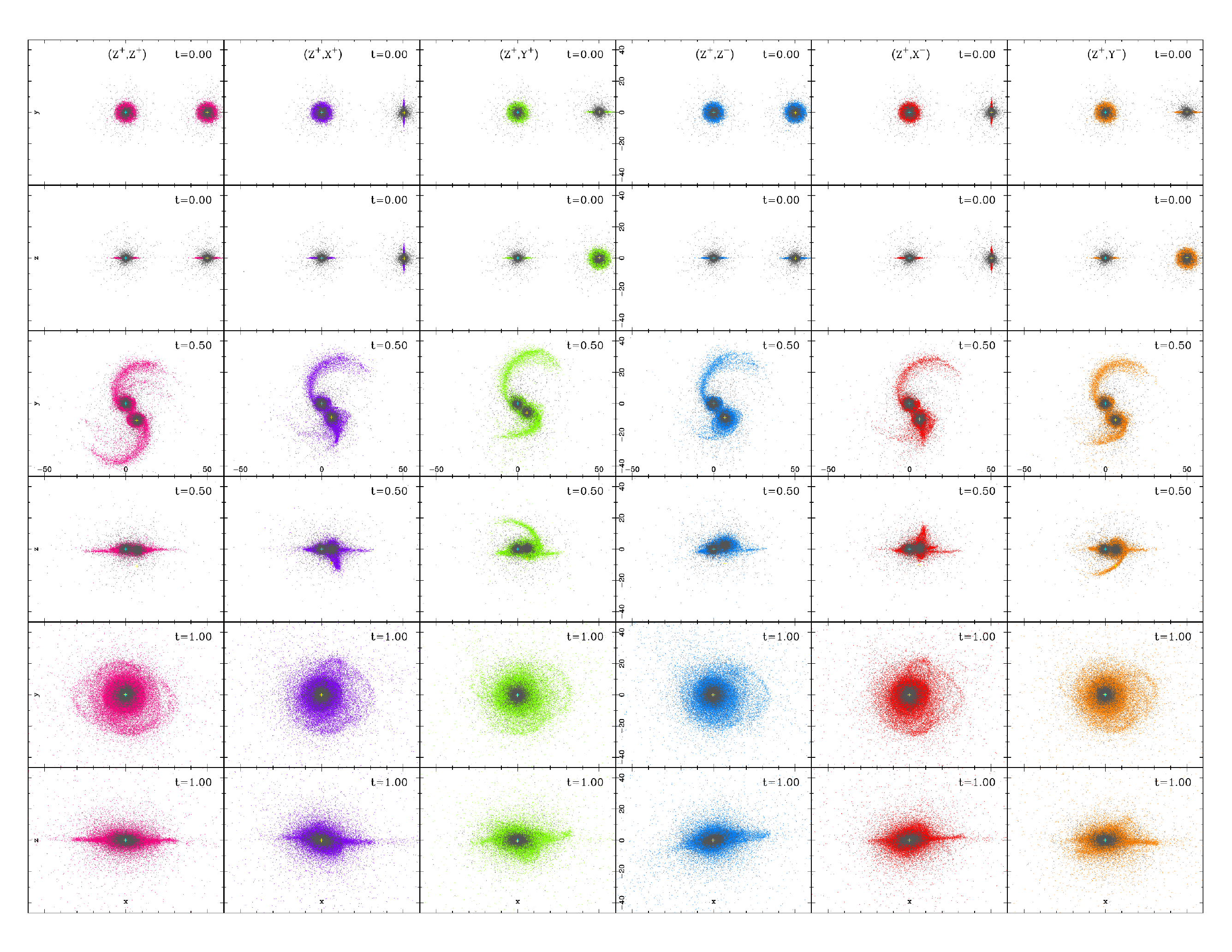}
	\caption{\small{ Bulge stars in grey and disk stars coloured, for all interactions of mass ratio 1:1 in which the rotaion of the primary galaxy is $\mbox{Z}^+$, a mode by column (color). The initial configuration at $t = 0.0$ Gyr (top), $t=0.5$ Gyr in middle, $t=1.0$ Gyr in the bottom. Each time has two perspectives, x-y plane and x-z plane, one above the other.}}
	\includegraphics[page=2, height=0.45\textheight]{same}
	\caption{\small{As Fig. 9  for all interactions with mass ratio 1:1  in  which the rotation  of primary galaxy is  $\mbox{X}^+$}.}
\end{figure}
\begin{figure}[H]
	\centering
	\includegraphics[page=3, height=0.45\textheight]{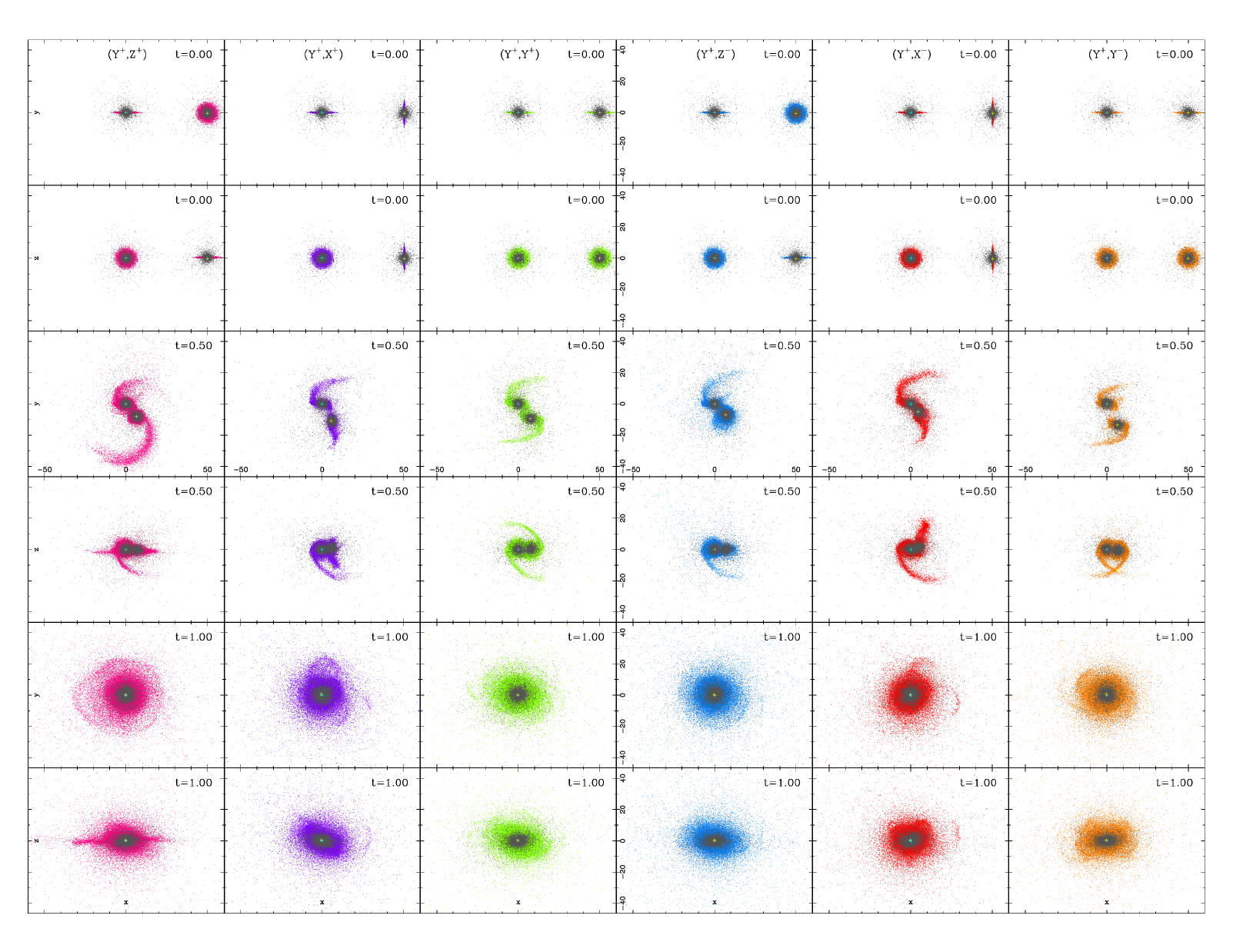}
	\caption{\small{As Fig. 9  for all interactions with mass ratio 1:1  in  which the rotation  of primary galaxy is $\mbox{Y}^+$.}}
	\includegraphics[page=4, height=0.45\textheight]{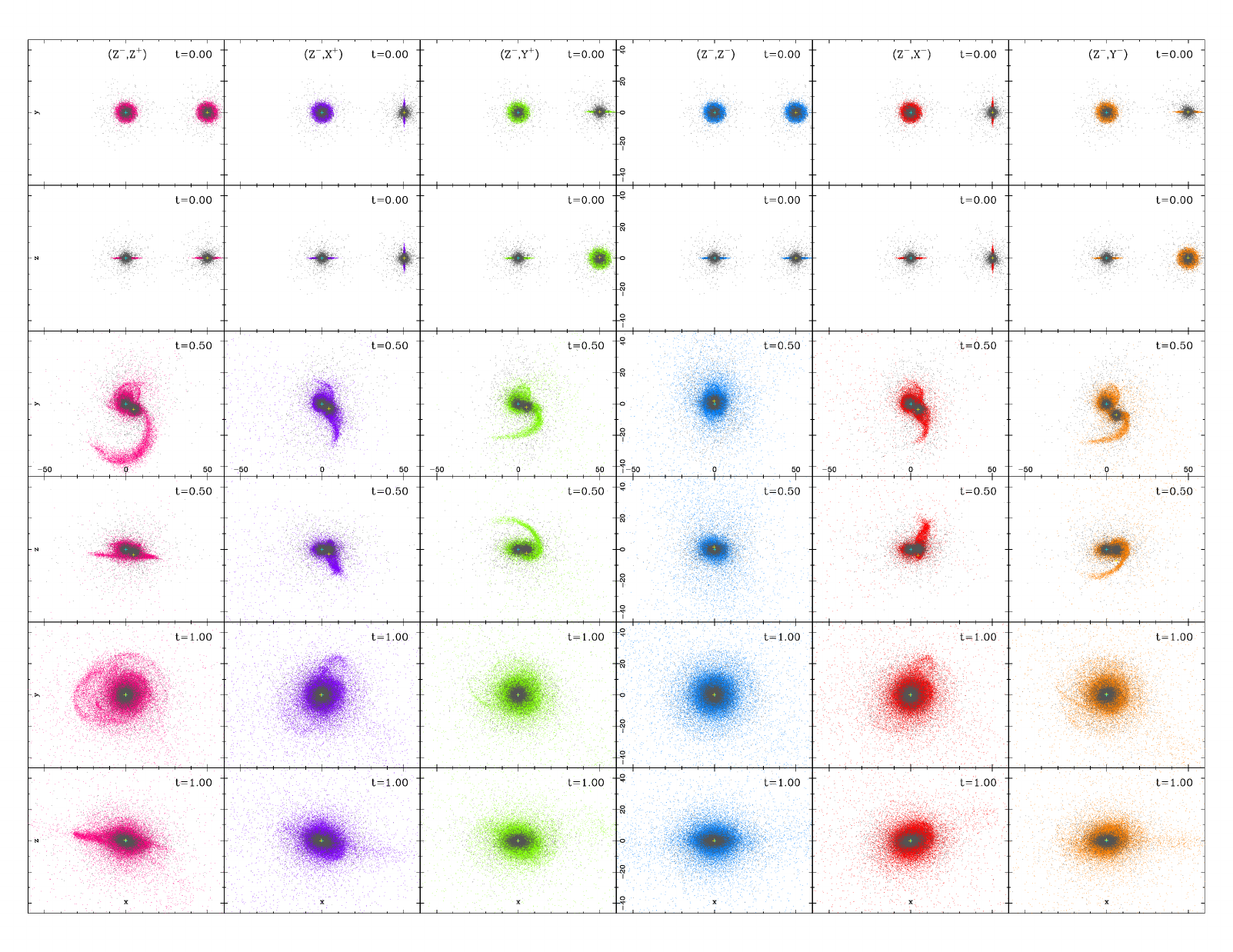}
	\caption{\small{ As Fig. 9  for all interactions with mass ratio 1:1  in  which the rotation  of primary galaxy is $\mbox{Z}^-$.}}
\end{figure}
\begin{figure}[H]
	\centering
	\includegraphics[page=5, height=0.45\textheight]{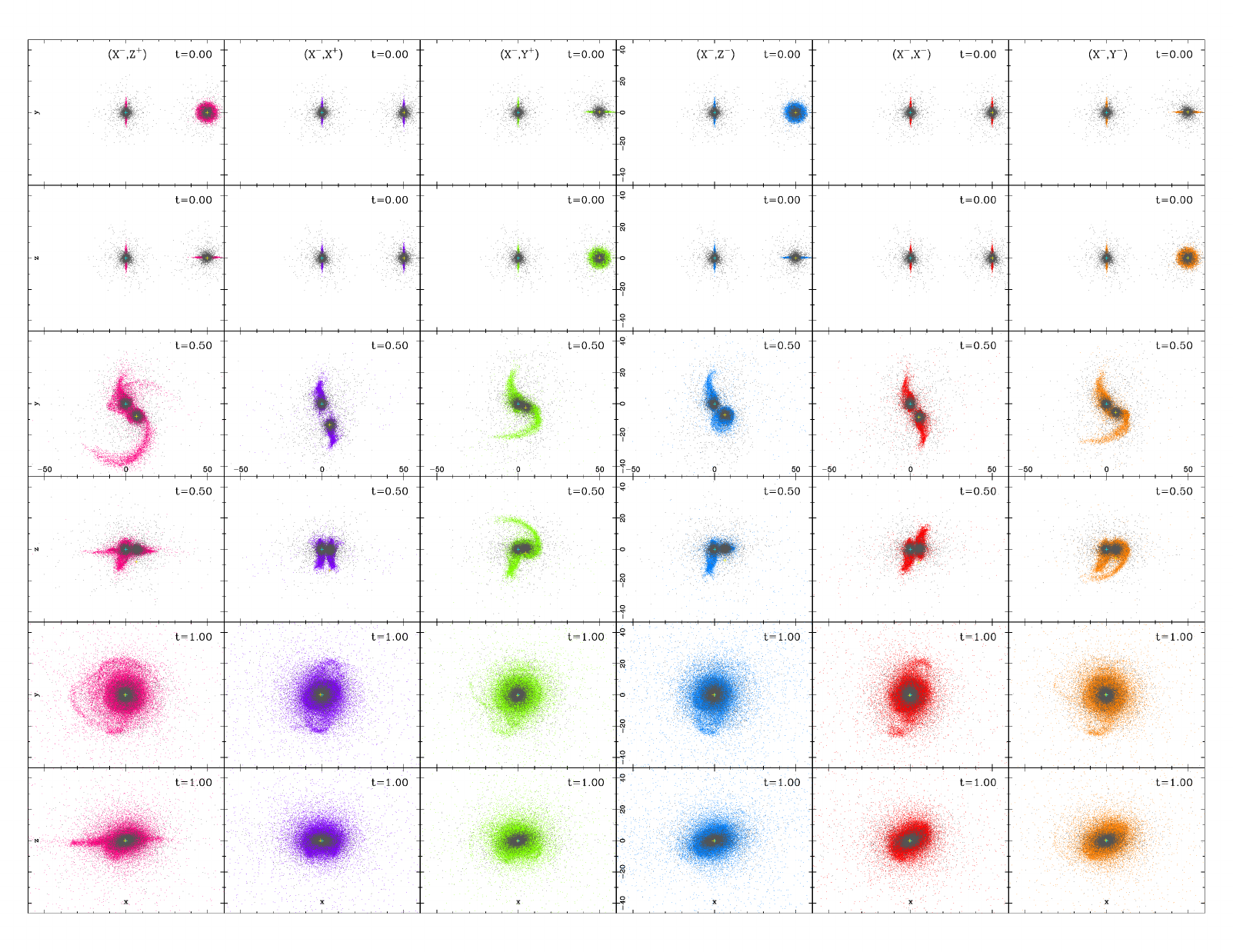}
	\caption{\small{As Fig. 9  for all interactions with mass ratio 1:1  in  which the rotation  of primary galaxy is $\mbox{X}^-$.}}
	\includegraphics[page=6, height=0.45\textheight]{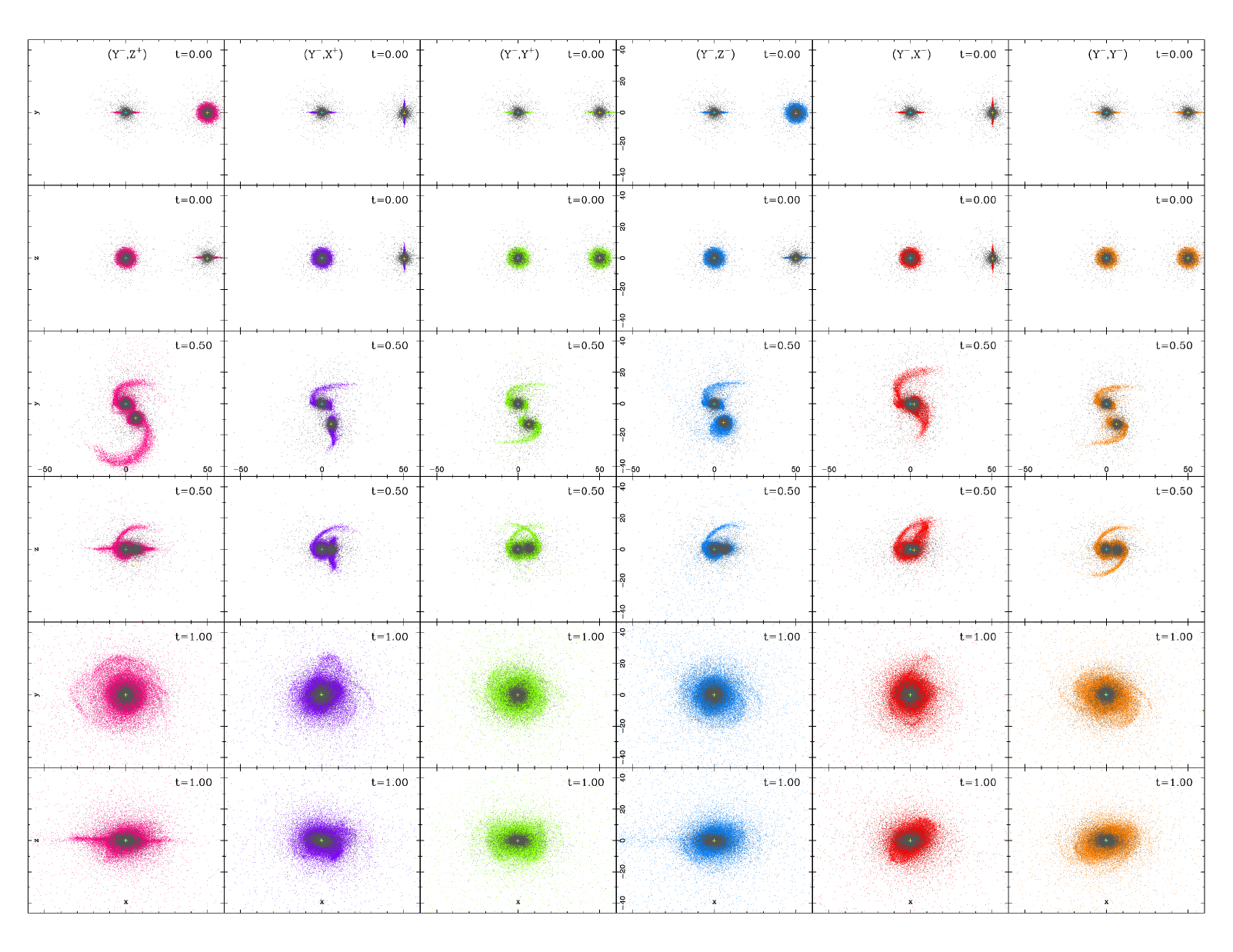}
	\caption{\small{ As Fig. 9  for all interactions with mass ratio 1:1  in  which the rotation  of primary galaxy is $\mbox{Y}^-$.}}
\end{figure}

\begin{figure}[H]
	\centering
	\subfigure{\includegraphics[page=1,width=0.48\textwidth]{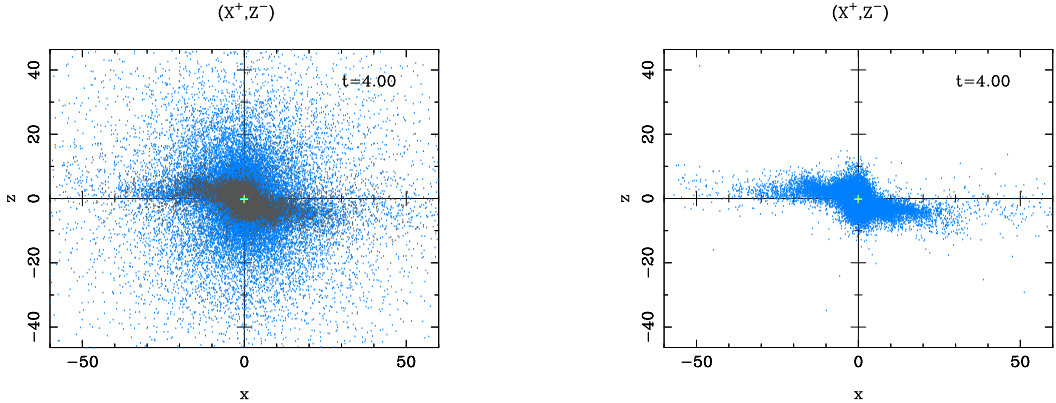}}
	\subfigure{\includegraphics[page=2,width=0.48\textwidth]{compa}}
	\caption{$(\mbox{X}^+,\mbox{Z}^-)$ interaction for 1:10 ratio at time $t=4.0$, the left side shows all galaxies stars and the right side the disks stars of the galaxies. The bulge stars are grey and the star and gas stars are blue. }
\end{figure}
\subsection{Comparison}

Table 1 shows the merging time ($\tau$). In addition and  at the time $t_f=4$ (0.98 h$^{-1}$Gyr, final time of the simulations), it is shown  the stellar mass $M_*$ (calculated from the mass $\mathcal{M}(R_d,z_0)$ of a cylinder of  radius $R_d$),  the scale length of the disk $R_d$ (calculated using the cylinder method  explained below), and the scale height $z_0$.  Values of $M_*(0)$, $R_d(0)$ and $z_0(0)$ are initial parameters of the primary disk ($G_1$ at time $t=0$),  The proportions in columns 4, 5 and 6 indicate how much the 
disk  is  taller, more extended or more  massive, compared with the initial disk of the primary galaxy. In each column the number on the left is the result for the mass ratio 1:10 and the number on the right is the result for the mass ratio 1:1, shown together for ease of comparison. \\Ê\\
\noindent
\textbf{Cylinder method}; For all modes of 1:10 and most cases of 1:1, (no matter orientation nor rotation) we see a new bigger disk in the  orbit plane (X-Y) because the orbital momentum dominates. This \textit{`new disk'} is analysed using the volume integration of equation (\ref{dendisk}). This is,
\begin{equation}
\mathcal{M} (R,z)=M_* \tanh \left( \dfrac{z}{z_0} \right) \left[1-\left(1+\dfrac{R}{R_d} \right)\exp\left( \dfrac{R}{R_d}\right) \right],
\label{masc}
\end{equation}
where $\mathcal{M} (R,z)$ is the mass inside a cylinder of radius $R$ and height $2z$ (the height of the cylinder is in the Z-direction for all the 36 cases for mass ratios  1:10 and 1:1). To determine the values of $R_d,~z_0$ and $M_*$ at time $t$, we measure the mass of two cylinders of the same height $z_r$ and different radii $r_1$ and $r_2$ ($r_1<r_2$) at time $t$, using equation (\ref{masc}) to get the fractional mass $f_r=\mathcal{M}(R_1,z_r)/\mathcal{M}(R_2,z_r)$ that satisfies
\begin{equation}
f_r \left[1-\left(1+\dfrac{R_2}{R_d} \right)\exp\left( \dfrac{R_2}{R_d}\right) \right]=1-\left(1+\dfrac{R_1}{R_d} \right)\exp\left( \dfrac{R_1}{R_d}\right).
\label{eqfr}
\end{equation} 
Then, we measure the mass of two cylinders of the same radius $R_h$ and different heights $z_1$ and $z_2$ ($z_1<z_2$) at time $t$, using equation (\ref{masc}) to get the fractional mass $f_h=\mathcal{M}(R_h,z_1)/\mathcal{M}(R_h,z_2)$ that satisfies
\begin{equation}
f_h \tanh \left( \dfrac{z_2}{z_0} \right)=\tanh \left( \dfrac{z_1}{z_0} \right).
\label{eqfh}
\end{equation}
It is necessary to explain that the cylinders used in this method have their centres in the primary bulge's centre of mass at the time $t$ (after the fusion). The equations (\ref{eqfr}) and (\ref{eqfh}) were solved numerically to obtain $R_d$ and $z_0$ at time $t$. From equation (\ref{masc}) for $R=R_d$ and $z_0=z_0$ we have
\begin{equation}
\mathcal{M} (R_d,z_0)=\dfrac{M_*}{(1-2e^{-1})\tanh(1)},
\end{equation}
Measuring the mass inside the cylinder $\mathcal{M}(R_d,z_0)$ we can calculate the  stellar mass $M_*$. 

\begin{table}[H]
	\centering
	\begin{tabular}{||c|| c c| c c| c c| c c| c c|} 
		\hline
	$ $ & \multicolumn{2}{c|}{} & \multicolumn{2}{c|}{} & \multicolumn{2}{c|}{} & \multicolumn{2}{c|}{} & \multicolumn{2}{c|}{} \\
    \textbf{Mode}    & 
	\multicolumn{2}{c|}{\textbf{$\tau$ (0.98 h$^{-1}$Gyr)}} & 
	\multicolumn{2}{c|}{\textbf{$M_*$ ($10^{10}$h$^{-1}$M$_\odot$)}}&
	\multicolumn{2}{c|}{\textbf{$M_*(t_f)/M_*(0)$}}&
	\multicolumn{2}{c|}{\textbf{$R_d(t_f)/R_d(0)$}} &  
	\multicolumn{2}{c|}{\textbf{$z_0(t_f)/z_0(0)$}}\\ 	  
		\hline \hline
$($$\mbox{Z}^{+}$$,$$\mbox{Z}^{+}$$)$ & 2.50 & 0.68 & 1.35 & 1.23 & 0.94 & 0.85 & 1.17 & 1.85 & 2.03 & 1.94 \\ 
$($$\mbox{Z}^{+}$$,$$\mbox{X}^{+}$$)$ & 2.45 & 0.70 & 1.20 & 1.13 & 0.83 & 0.78 & 1.15 & 1.51 & 2.02 & 2.05 \\ 
$($$\mbox{Z}^{+}$$,$$\mbox{Y}^{+}$$)$ & 2.42 & 0.67 & 0.98 & 1.00 & 0.68 & 0.69 & 1.13 & 1.42 & 1.57 & 1.98 \\ 
$($$\mbox{Z}^{+}$$,$$\mbox{Z}^{-}$$)$ & 2.43 & 0.75 & 1.18 & 1.16 & 0.82 & 0.80 & 1.19 & 1.46 & 1.71 & 2.19 \\ 
$($$\mbox{Z}^{+}$$,$$\mbox{X}^{-}$$)$ & 2.38 & 0.73 & 1.33 & 0.98 & 0.92 & 0.68 & 1.21 & 1.45 & 1.98 & 1.80 \\ 
$($$\mbox{Z}^{+}$$,$$\mbox{Y}^{-}$$)$ & 2.53 & 0.68 & 1.00 & 1.02 & 0.69 & 0.70 & 1.12 & 1.50 & 1.47 & 1.83 \\
		\hline 
$($$\mbox{X}^{+}$$,$$\mbox{Z}^{+}$$)$ & 3.26 & 0.68 & 0.48 & 1.33 & 0.33 & 0.92 & 0.92 & 1.47 & 1.99 & 2.32 \\ 
$($$\mbox{X}^{+}$$,$$\mbox{X}^{+}$$)$ & 3.08 & 0.66 & 0.56 & 0.99 & 0.39 & 0.68 & 0.96 & 1.34 & 2.35 & 1.98 \\ 
$($$\mbox{X}^{+}$$,$$\mbox{Y}^{+}$$)$ & 3.15 & 0.63 & 0.56 & 0.80 & 0.39 & 0.56 & 1.03 & 1.31 & 2.07 & 2.01 \\ 
$($$\mbox{X}^{+}$$,$$\mbox{Z}^{-}$$)$ & 3.02 & 0.63 & 0.62 & 1.06 & 0.43 & 0.73 & 0.95 & 1.32 & 2.61 & 2.04 \\ 
$($$\mbox{X}^{+}$$,$$\mbox{X}^{-}$$)$ & 2.95 & 0.72 & 0.54 & 0.79 & 0.38 & 0.55 & 0.96 & 1.19 & 2.20 & 2.31 \\ 
$($$\mbox{X}^{+}$$,$$\mbox{Y}^{-}$$)$ & 3.00 & 0.64 & 0.46 & 0.82 & 0.32 & 0.56 & 0.97 & 1.26 & 1.79 & 1.75 \\ 
		\hline
$($$\mbox{Y}^{+}$$,$$\mbox{Z}^{+}$$)$ & 3.22 & 0.65 & 0.67 & 1.15 & 0.47 & 0.79 & 0.95 & 1.50 & 2.23 & 2.26 \\ 
$($$\mbox{Y}^{+}$$,$$\mbox{X}^{+}$$)$ & 2.97 & 0.74 & 0.42 & 0.78 & 0.29 & 0.54 & 0.94 & 1.29 & 1.71 & 2.19 \\ 
$($$\mbox{Y}^{+}$$,$$\mbox{Y}^{+}$$)$ & 2.88 & 0.71 & 0.43 & 1.18 & 0.30 & 0.82 & 0.93 & 1.46 & 2.02 & 2.52 \\ 
$($$\mbox{Y}^{+}$$,$$\mbox{Z}^{-}$$)$ & 2.96 & 0.74 & 0.48 & 1.04 & 0.33 & 0.72 & 0.98 & 1.31 & 1.99 & 2.22 \\ 
$($$\mbox{Y}^{+}$$,$$\mbox{X}^{-}$$)$ & 2.89 & 0.69 & 0.58 & 0.70 & 0.40 & 0.48 & 0.98 & 1.29 & 2.28 & 2.15 \\ 
$($$\mbox{Y}^{+}$$,$$\mbox{Y}^{-}$$)$ & 2.92 & 0.78 & 0.50 & 0.68 & 0.35 & 0.47 & 0.90 & 1.32 & 2.05 & 2.19 \\ 
		\hline
$($$\mbox{Z}^{-}$$,$$\mbox{Z}^{+}$$)$ & 3.14 & 0.64 & 0.74 & 0.89 & 0.51 & 0.61 & 0.97 & 1.39 & 1.64 & 2.02 \\ 
$($$\mbox{Z}^{-}$$,$$\mbox{X}^{+}$$)$ & 3.13 & 0.65 & 0.74 & 0.74 & 0.51 & 0.51 & 0.99 & 1.33 & 1.57 & 1.75 \\ 
$($$\mbox{Z}^{-}$$,$$\mbox{Y}^{+}$$)$ & 3.00 & 0.69 & 0.80 & 0.87 & 0.55 & 0.60 & 1.01 & 1.31 & 1.75 & 2.20 \\ 
$($$\mbox{Z}^{-}$$,$$\mbox{Z}^{-}$$)$ & 3.02 & 0.66 & 0.77 & 1.16 & 0.53 & 0.80 & 1.02 & 1.37 & 1.69 & 2.18 \\ 
$($$\mbox{Z}^{-}$$,$$\mbox{X}^{-}$$)$ & 3.07 & 0.69 & 0.74 & 1.18 & 0.51 & 0.82 & 0.97 & 1.29 & 1.60 & 2.76 \\ 
$($$\mbox{Z}^{-}$$,$$\mbox{Y}^{-}$$)$ & 3.00 & 0.69 & 0.82 & 0.83 & 0.57 & 0.58 & 0.98 & 1.42 & 1.55 & 1.80 \\ 
		\hline
$($$\mbox{X}^{-}$$,$$\mbox{Z}^{+}$$)$ & 3.17 & 0.65 & 0.45 & 1.07 & 0.31 & 0.74 & 0.96 & 1.45 & 1.80 & 2.02 \\ 
$($$\mbox{X}^{-}$$,$$\mbox{X}^{+}$$)$ & 3.06 & 0.71 & 0.70 & 0.94 & 0.48 & 0.65 & 1.01 & 1.26 & 2.50 & 2.42 \\ 
$($$\mbox{X}^{-}$$,$$\mbox{Y}^{+}$$)$ & 3.11 & 0.66 & 0.40 & 0.74 & 0.28 & 0.51 & 0.92 & 1.25 & 1.87 & 1.92 \\ 
$($$\mbox{X}^{-}$$,$$\mbox{Z}^{-}$$)$ & 3.08 & 0.73 & 0.67 & 1.08 & 0.46 & 0.74 & 0.97 & 1.26 & 1.96 & 2.38 \\ 
$($$\mbox{X}^{-}$$,$$\mbox{X}^{-}$$)$ & 3.04 & 0.70 & 0.62 & 0.90 & 0.43 & 0.62 & 0.99 & 1.36 & 2.07 & 2.01 \\ 
$($$\mbox{X}^{-}$$,$$\mbox{Y}^{-}$$)$ & 3.10 & 0.63 & 0.56 & 0.80 & 0.38 & 0.56 & 0.93 & 1.28 & 2.21 & 2.00 \\ 
		\hline
$($$\mbox{Y}^{-}$$,$$\mbox{Z}^{+}$$)$ & 2.82 & 0.69 & 0.54 & 1.17 & 0.38 & 0.81 & 0.86 & 1.44 & 2.08 & 1.87 \\ 
$($$\mbox{Y}^{-}$$,$$\mbox{X}^{+}$$)$ & 2.82 & 0.76 & 0.63 & 0.99 & 0.44 & 0.69 & 0.86 & 1.24 & 2.26 & 2.02 \\ 
$($$\mbox{Y}^{-}$$,$$\mbox{Y}^{+}$$)$ & 2.81 & 0.74 & 0.51 & 1.33 & 0.35 & 0.92 & 0.87 & 1.24 & 2.16 & 2.83 \\ 
$($$\mbox{Y}^{-}$$,$$\mbox{Z}^{-}$$)$ & 2.73 & 0.71 & 0.65 & 1.06 & 0.45 & 0.73 & 0.82 & 1.26 & 2.36 & 1.84 \\ 
$($$\mbox{Y}^{-}$$,$$\mbox{X}^{-}$$)$ & 2.72 & 0.64 & 0.47 & 0.96 & 0.32 & 0.67 & 0.96 & 1.31 & 1.76 & 2.00 \\ 
$($$\mbox{Y}^{-}$$,$$\mbox{Y}^{-}$$)$ & 2.74 & 0.72 & 0.44 & 1.00 & 0.31 & 0.69 & 0.87 & 1.38 & 1.80 & 1.86 \\
		\hline
	\end{tabular}
	\caption{Chosen parameters to characterize  the resulting galaxy morphology of the two galaxies interactions with different orientations and rotations, The number on the left of each column has mass ratio 1:10 and the number on the right is for mass ratio 1:1.}
	\label{table:1}
\end{table}
\newpage\noindent

        \begin{figure}[H]
        \centering
        \includegraphics[width=\textwidth,height=7.5cm ]{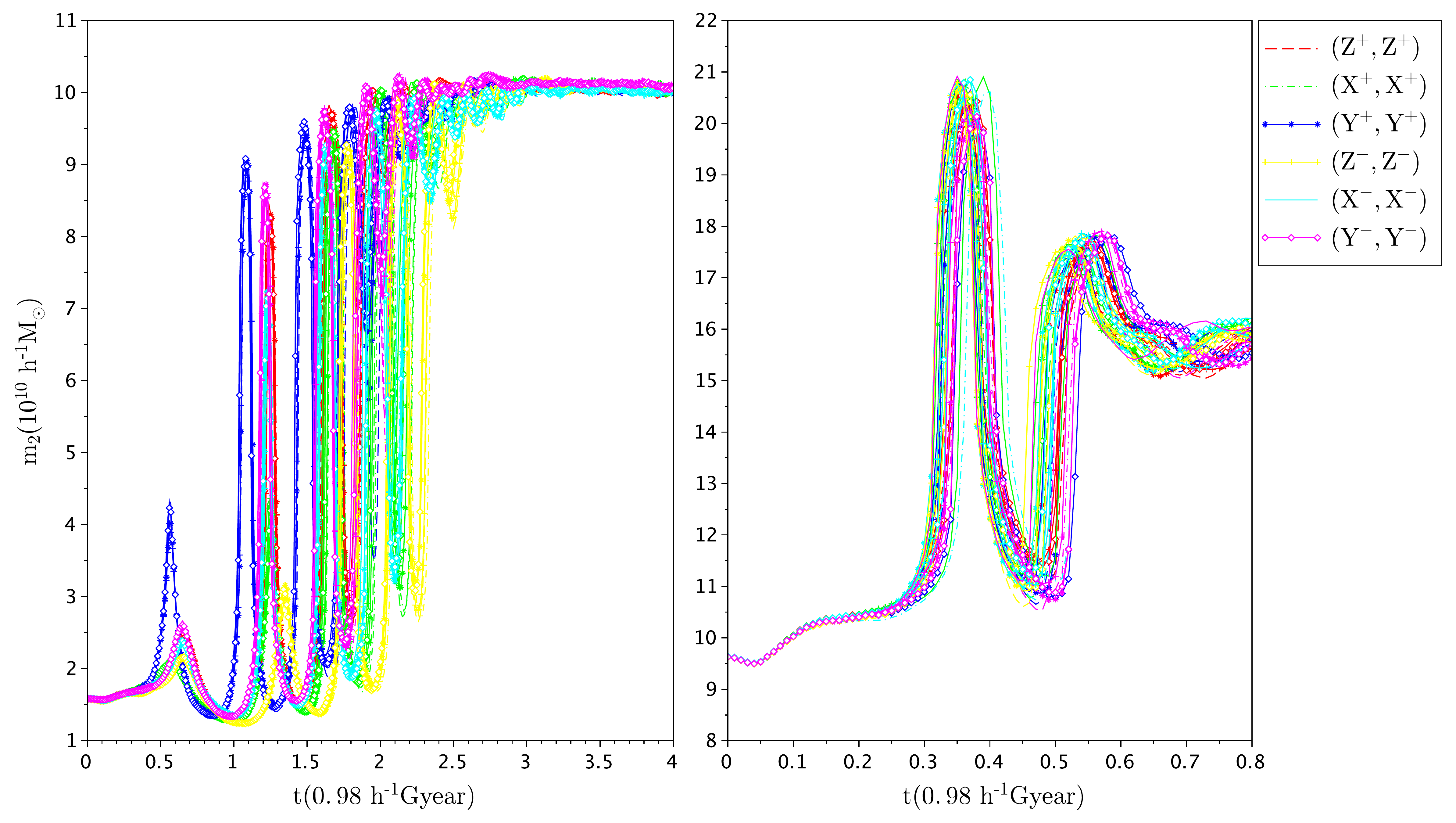}
        \caption{Comparison between  mass ratios 1:10 (left) and 1:1 (right)  of the evolution of the mass of the secondary galaxy, $M_2$.}
        \end{figure}
Figure 17 shows the evolution of the mass of the secondary galaxy. For the 1:10 cases the galaxy increases its mass steadily until the merger product includes most of the mass of the primary galaxy as well. In the 1:1 mass ratio cases there is a notable contraction of both galaxies visible in the simulation sat $t \sim 0.35$ Gyr which brings greater mass within the radius used to measure the mass of the secondary galaxy.Then, the mass  is reduced as the simulation progresses and the mass moves further away from the centre again. As one could expected, the dynamics of two equal galaxies evolves much faster than the evolution with two galaxies quite different in masses.
%
 \section{Discussions and conclusions}
We have used  the GADGET-2 N-body code to study  the galaxy morphology and the time taken by two interacting galaxies (mass ratios 1:10 and 1:1 with the same bulge-to-disk ratios) to merge due to gravity. 
It was found that for case with a 1:10 mass ratio some interaction modes are grouped, with basically the same behavior for each group.  
These groups do not exist for case with galaxies of 1:1 mass ratio, where there is a unique group, and every case has almost the same evolution of distances between bulges and about the same merging time.\\
 \\
The evolution of the masses of each galaxy is completely different in both cases. For the case with a 1:10 mass ratio, the global maximum is located at the end evolution, meaning that the second galaxy increases its mass constantly. For the case  with mass ratio 1:1, the global maximum is located around $t=0.35$ Gyr, causing a reduction of the merging time. Of course, the galaxy morphology vary for each case due to the difference on masses since the dispersion of stars for the case 1:10 is bigger than the dispersion of stars in the case 1:1, for most of the interaction types. 
\\    
\\
The evolution of the galaxies is principally affected by the angular momentum of interaction of both galaxies, and secondly by rotation of the more massive galaxy.  
For a 1:10 mass ratio, the rotation can reduce or increase the merging time by an interval of  $\sim0.9$ (0.98 h$^{-1}$Gyr). In contrast, for the 1:1 mass ratio case, the relative rotation of galaxies   can reduce or increase the merging time by an interval of $\sim0.2$ (0.98 h$^{-1}$Gyr). \\
\\

In all cases with 1:10 mass ratios and cases with the secondary galaxy with rotation and orientation $\mbox{Z}^{+}$ the merger product contains a disk in the $x-y$ plane, usually greatly extended with respect to the original primary disk both in radius and height, due to the high initial angular momentum about the $z$ axis. 
The 1:10 mass ratio cases all take a long time to complete merging and produce a disk-like galaxy in all cases, in the plane of the original secondary orbit due to the high initial angular momentum in this plane. the disk is usually extended and distorted. During the interaction shell-like structures are formed in the plane of the final disk. though their formation from a companion with a tangential orbit contradicts the results of \cite{kar}.\\Ê\\

For the cases with equal masses the merging is more rapid and the disks less extended usually and the dstructures at intermediate times look more like parts of spiral arms. When the secondary galaxy is in the rotation state $\mbox{Z}^{+}$ the final product has a disturbed, asymmetric, warped disk, while $\mbox{Z}^{+}$,$\mbox{Z}^{+}$ has a box-peanut-like bulge. The other cases end in a  bulge/ellipsoid-dominated galaxy (some box-peanut-like) due to the high initial angular momentum perpendicular to the $x-y$ plane. 

\vskip1pc\noindent
\section*{Acknowledgements}
We want to thank Dr. Iv\^{a}nio Puerari and   M.C. Diego Valencia Enr\'{\i}quez from INAOE institute in Tonantzintla, Puebla, M\'exico for their help  teaching to one of us how to use GADGET-2 code. 
\vskip1pc\noindent

\bibliographystyle{aasjournal}
\bibliography{biblio2short} 
\end{document}